\documentclass[aps,preprint,amsmath,amssymb]{revtex4}
\usepackage{graphicx, color}

\def\be{\begin{eqnarray}}
\def\ed{\end{eqnarray}}
\def\non{\nonumber}
\def\ga{\gamma}
\def\e{\epsilon}
\def\ve{\varepsilon}

\begin{document}
\title{\Large \bf  Impact of colored scalars on $D^0-\bar D^0$ mixing in diquark
models }

\date{\today}

\author{ \bf  Chuan-Hung Chen$^{1,2}$\footnote{Email:
physchen@mail.ncku.edu.tw}
 }

\affiliation{ $^{1}$Department of Physics, National Cheng-Kung
University, Tainan 701, Taiwan \\
$^{2}$National Center for Theoretical Sciences, Hsinchu 300, Taiwan
 }

\begin{abstract}
Inspired by the recent observation of the $D^0-\bar D^0$ mixing, we
explore the effects of colored scalars on the $\Delta C=2$ process
in diquark models. As an illustration, we investigate the diquarks
with the quantum numbers of  (6, 1, 1/3) and (6, 1, 4/3) under
$SU(3)_C\times SU(2)_L\times U(1)_Y$ gauge symmetries, which
contribute to the process at one-loop and tree levels, respectively.
We show that $\Delta m_D$ gives the strongest constraint on the free
parameters. In addition, we find that the small couplings can be
naturally interpreted by the suppressed flavor mixings if the
diquark of (6, 1, 4/3) only couples to the third generation.

\end{abstract}
\maketitle

Unlike $K$ and $B$ systems,  the short-distance (SD) contributions
to charmed-meson flavor changing neutral current (FCNC) processes,
such as the $D^0-\bar D^0$ mixing \cite{DD_SM} and the decays of
$c\to u \ell^{+} \ell^{-}$ and $D\to \ell^{+} \ell^{-}$
\cite{BGHP_PRD52}, are highly suppressed due to the stronger
Glashow-Iliopoulos-Maiani (GIM) mechanism \cite{GIM} and weaker
heavy quark mass enhancements in the loops. In addition, it is often
claimed that the long-distance (LD) effect for the $D^0-\bar D^0$
mixing should be the prevailing contribution in the SM.
Nevertheless, because the nonperturbative hadronic effects are hard
to control, the issue of the LD dominance is still inconclusive
\cite{order3,Petrov,order2-1,order2-2}. Recently, besides the
progress of observing the $B_s$ oscillation
\cite{CDF,D0,BsCP,CG_PRD76,CGL_PLB670}, the evidence for the
$D^0-\bar D^0$ mixing has also been exposed with the world averages
\cite{HFAG}
\be \label{D2}
x_D&\equiv& {\Delta m_{D}\over \Gamma_{D}}\;=\; (0.811\pm0.334)\%\,,\non\\
y_D&\equiv& {\Delta \Gamma_{D}\over 2\Gamma_{D}}\;=\;(0.309\pm0.281)\%\,,\non\\
  y_{DCP}&=&(1.072\pm 0.257)\% \,,
 \ed
where $x_D (y_D)$ denotes the mass (lifetime) difference parameter
and $y_{DCP}$ is the mixing parameter including the CP violating
information. That is, if no CP violation is found in the $D$-meson
oscillation, we have $y_{DCP}=y_D$. Due to these data, lots
of studies on the physics beyond Standard Model (SM) have been done
\cite{Nir,Combined,GPP,New_D,Blanke,LRM,CGY_PLB655}.
In particular, the possible extensions of the SM for the $D^0-\bar D^0$ mixing have
been investigated by the authors in Ref.~\cite{GHPP} in great
detail.

In this note, we explore the issue in scalar diquark models which
were not included in the previous discussions \cite{GHPP}.
In the literature, the motivation to study the light colored scalar
could be traced to the solution for the strong CP problem
\cite{BZ_PRL55}, where to avoid the domain-wall problem on the
spontaneous CP violating mechanism, the models were constructed in
the framework of grand unified theories (GUTs), e.g. SU(5) gauge
symmetry. The associated new source of CP violation on K and B
systems was also studied in Refs.~\cite{Barr_PRD34,BF_PRD41}. In
addition, since the scalar sector in the SM has not been tested
experimentally, it is plausible to assume the existence of other
possible scalars in the gauge symmetry of $SU(3)_{C} \times
SU(2)_{L} \times U(1)_Y$. Accordingly, the general scalar
representations could be ${\bf( 1, 2)}_{1/2}$, ${\bf (8, 2)}_{1/2}$,
${\bf (6, 3)}_{1/3}$, ${\bf (6, 1)}_{4/3,1/3,-2/3}$, ${\bf(3,
3)}_{-1/3}$, ${\bf(3, 1)}_{2/3,-1/3,-4/3}$, where the first (second)
argument in the brackets denotes the representation in color (weak
isospin) space and the number in the subscript corresponds to the
hypercharge of $U(1)_Y$ \cite{MW}. Besides the SM Higgs doublet, it
has been shown in Ref.~\cite{MW} that when the hypothesis of minimal
flavor violation (MFV) is imposed, only the representation ${\bf (8,
2)}_{1/2}$ could avoid FCNCs at tree level. As a result, due to the
suppression of Cabibbo-Kobayashi-Maskawa (CKM) matrix elements and
the masses of light quarks, the loop induced $D^0-\bar D^0$
oscillation in the color octet model is also negligible. Therefore,
in the following analysis, we will concentrate on the situation of
color triplet and sextet.
In terms of involved Feynman diagrams, we find that $\Delta m_D$ can
be produced by the diquark models through both box and tree
diagrams. The various possible scalar diquarks are presented in
Table~\ref{tab:diquark}, where the second column in the table
denotes the representations of the diquarks under $SU(3)_C\times
SU(2)_L \times U(1)_Y$,
the third column gives the interactions of quarks and diquarks, the
fourth column displays the relation of couplings in flavors and the
last column shows the type of the effect that generates $\Delta
m_D$. From the table, we see that only Model (7) can lead to the
$\Delta C=2$ interaction at tree level. Due to the antisymmetric
property in flavor indices, Model (6) cannot contribute to the
$\Delta C=2$ process at  tree level. It is worth mentioning that the
colored sextet scalar also exists in a class of partial unification
theories based on $SU(2)_L \times SU(2)_R \times SU(4)_C$
\cite{MOY_PRD77}.

\begin{table}[hptb]
\caption{Various diquark models for the $D^0-\bar D^0$ mixing with
$\e = i\tau_2$. }\label{tab:diquark}
\begin{ruledtabular}
\begin{tabular}{ccccc}
Model & H  & Interaction & flavor symmetry & diagram
 \\ \hline
(1) & $(3,1,-1/3)$ & $f_{ij} \bar Q^c_{i\alpha} \e Q_{j\beta}\ve^{\alpha\beta\ga} H_{\ga}$ & $f_{ij}=f_{ji}$ & box \\
(2) & $(3,1,-1/3)$ & $f_{ij} \bar d_{i\alpha} u^c_{j\beta}\ve^{\alpha\beta\ga} H^\dagger_{\ga}$ & $--$ & box \\
(3) & $(6,1,1/3)$ & $f_{ij} \bar Q^c_{i\alpha} \e Q_{j\beta} H^{\dagger\alpha\beta}$ & $f_{ij}=-f_{ji}$ & box \\
(4) & $(6,1,1/3)$ & $f_{ij} \bar d_{i\alpha} u^c_{j\beta} H^{\alpha\beta}$ & $--$ & box \\
(5) & $(3,3,-1/3)$ & $f_{ij} \bar Q^c_{i\alpha} \e H_{\ga} Q_{j\beta}\ve^{\alpha\beta\ga} $ & $f_{ij}=-f_{ji}$ & box \\
(6) & $(3,1,-4/3)$ & $f_{ij} \bar u_{i\alpha}  u^c_{j\beta}\ve^{\alpha\beta\ga} H^{\dagger}_{\ga}$ & $f_{ij}=-f_{ji}$ & box \\
(7) & $(6,1,4/3)$ & $f_{ij} \bar u_{i\alpha}  u^c_{j\beta} H^{\alpha\beta}$ & $f_{ij}=f_{ji}$ & tree \\
\end{tabular}
\end{ruledtabular}
\end{table}

Since our purpose is to show the influence of diquarks on the $D^0-\bar
D^0$ oscillation, we are not planning to calculate the contributions
of each model
shown in Table~\ref{tab:diquark}. For
comparison, we use Models (4) and (7), which have the same color
structure, to illustrate the diquark effects.
It is expected that the contributions of other models should be
similar in order of magnitude.

\underline {1. \textit{diquark (6,1,1/3)} }

\begin{figure}[htbp]
\includegraphics*[width=5.0 in]{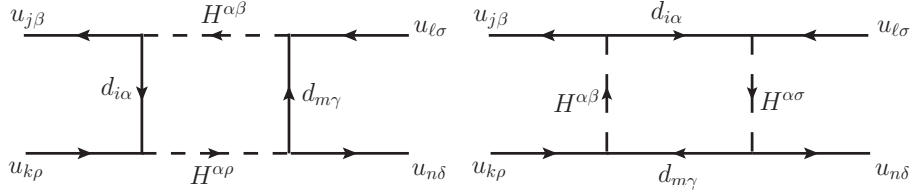}
\caption{Diquark box diagrams for the $D^0-\bar D^0$ mixing.}
 \label{fig:H6_loop}
\end{figure}

We first write the interaction of quarks and the diquark of (6,1,1/3)
as
 \be
{\cal L}_{H_{\bf 6}} = f_{ij} d^{c^T}_{i\alpha} C P_L u^c_{j\beta}
H^{\alpha\beta}_{\bf 6} +h.c. \,,\label{eq:LH6-1}
 \ed
where $f_{ij}$ denote the couplings of diquark and various flavors,
$C=i\ga^0 \ga^2$ is the charge-conjugation operator, $P_{L(R)}=(1\mp
\ga_5)/2$ is the chiral projection operator and
$H^{\alpha\beta}_{\bf 6}$ is a weak gauge-singlet and colored sextet
scalar with $\alpha$ and $\beta$ being the color indices. After
Fierz transformation, since the structure of four-fermion
interactions becomes $\bar d_i\Gamma d_j \bar c_m\Gamma c_{n}$,
obviously lifetime and mass differences in the $D^0-\bar D^0$ mixing
cannot be induced at tree level. However,
they can
be produced at one-loop where the box diagrams are sketched in
Fig.~\ref{fig:H6_loop}.

To formulate the $\Delta C=2$ effective Hamiltonian, we set the
flavor indices in Fig.~\ref{fig:H6_loop} to be $k=\ell=c$ and
$j=n=u$. Hence, by including Wick contractions and neglecting the
external momenta and the internal masses of light quarks, the
effective Hamiltonian for Fig.~\ref{fig:H6_loop} is written as
 \be
-i H_{\Delta C=2} &=& \frac{1}{2}C^2_D \left(5\delta_\beta^\sigma
\delta_\rho^\delta + \delta_\delta^\sigma \delta_\beta^\rho
\right)\int \frac{d^4 q}{(2\pi)^4} \frac{q^\mu q^\nu}{ (q^2)^2
(q^2-m^2_H)^2} \bar u_\beta \ga_\mu P_R c_\rho \bar u_\delta \ga_\mu
P_R c_\sigma \label{eq:2c}
 \ed
with $C_D = \sum_i f^*_{ic} f_{iu}$. Here, we have used the
propagator for the sextet scalar as
 \be
\langle T H_{\alpha\beta}H^{\ga\sigma} \rangle = \int \frac{d^4
q}{(2\pi)^4}\frac{ e^{-iq\cdot x}}{q^2-m^2_H+i\e}
\left(\delta^\ga_\alpha \delta^\sigma_\beta + \delta^\sigma_\alpha
\delta^\ga_\beta \right)\,. \label{eq:H6prop}
 \ed
With the loop integral
 \be
\int \frac{d^4 q}{(2\pi)^4} \frac{q^\mu q^\nu}{ (q^2)^2
(q^2-m^2_H)^2} = -i\frac{g^{\mu\nu}}{4(4\pi)^2 m^2_H}\,, \non
 \ed
Eq.~(\ref{eq:2c}) can be expressed as
 \be
{\cal H}_{\Delta C=2} &=& \frac{C^2_D}{128\pi^2 m^2_H} \left[ 5 \bar
u_\beta \ga_\mu P_R c_\rho \bar u_\rho \ga^\mu P_R c_\beta + \bar u
\ga_\mu P_R c \bar u \ga^\mu P_R c\right]\,.
  \ed
Using the transition matrix element given by
 \be
\langle \bar D^0| \bar u_R \ga_\mu  c_R \bar u_R \ga^\mu  c_R
|D^0\rangle= \frac{2}{3} f^2_D m^2_D B_D\,, \label{eq:FDD}
 \ed
the D mixing parameters are
 \be
 x_D &=&\frac{\Delta m_D}{\Gamma_D} = Z_D \left( \frac{C_D}{m_H}
\right)^2\,, \non\\
Z_D &=& \tau_D
\frac{19f^2_D m_D B_D}{768\pi^2  } \label{eq:deltamD}
 \ed
 with $\Delta m_D=2 |M_{12}|=2|\langle D | {\cal H}_{\Delta C=2}| D\rangle|$.
Taking $\tau_D=4.10\times 10^{-13} s$, $m_D=1.86$ GeV, $f_D=0.222$
GeV \cite{PDG08} and $B_D=0.82$ \cite{GBS_PRD55}, in order to fit
the current experimental data, the unknown parameters should satisfy
 \be
\left( \frac{C_D}{m_H} \right)^2 &=& \frac{ x_D}{Z_D}=4.7\times
10^{-7} x_D {\rm GeV^{-2}}\,. \label{eq:xD_limit}
 \ed
With $m_H\sim 1$ TeV and $x_D\sim 8\times 10^{-3}$, the
free parameter $|C_D| $ is about  $0.06$.

\begin{figure}[htbp]
\includegraphics*[width=2. in]{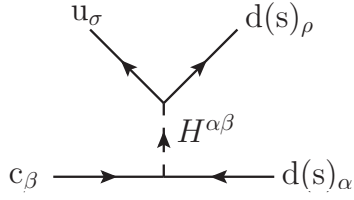}
\caption{Diquark-mediated flavor diagram for $D^+\to \pi^+ (\pi^0,
\phi)$.}
 \label{fig:D1}
\end{figure}

Besides a serious constraint on the free
parameters from
$\Delta m_D$, for comparison, we  consider other possible limits
from rare non-leptonic D decays, such as $D^+\to \pi^+ \pi^0 $ and
$D^+\to \pi^+ \phi$ decays,
which are Cabibbo-suppressed and
Cabibbo and color-suppressed processes in the
SM, respectively. Their current measurements are \cite{PDG08}
 \be
 B(D^+\to \pi^+ \pi^0 ) &=& (1.24\pm 0.07)\times 10^{-3}\,,\non\\
 B(D^+\to \pi^+ \phi) &=& (6.2\pm 0.7)\times 10^{-3}.
 \ed
According to the interactions in Eq.~(\ref{eq:LH6-1}), flavor
diagrams for D decays are given in Fig.~\ref{fig:D1} and the
corresponding interactions for $c\to u \bar d d (\bar s s)$
are found to be
 \be
 {\cal H}_{c\to u}&=& - \frac{f^*_{qc} f_{qu}}{2m^2_H}
 \left(O^{q}_{1}+O^{q}_{2} \right)\,,\non\\
 O^{q}_{1} &=& \bar u \ga^\mu P_R c \bar q \ga_\mu P_R c \,,\non \\
 O^{q}_{2} &=& \bar u_\alpha \ga^\mu P_R c_\beta \bar q_\beta \ga_\mu P_R c_\alpha
 \ed
with $q=d$ and $s$.
Based on the decay constants and transition form factor, defined by
 \be
\langle 0 | \bar q' \ga^\mu \ga_5 q|P(p)\rangle &=& if_P p^\mu\,,\ \
\ \langle 0 | \bar s \ga^\mu  s|\phi(p)\rangle = i m_\phi f_\phi
\varepsilon^\mu_{\phi}(p)\,,  \non\\
 \langle \pi(p_2)| \bar u \ga_\mu c| D (p_1)\rangle &=&
f_{+}(k^2)\Big\{P_{\mu}-\frac{P\cdot k }{k^2}k_{\mu} \Big\}
+\frac{P\cdot k}{k^2}f_{0}(k^2)\,k_{\mu}
 \ed
with $P=p_1+p_2$ and $k=p_1-p_2$, the decay amplitudes by the naive
factorization approach for $D^+\to \pi^+ (\pi^0,\phi)$ are given by
 \be
 A(\pi^+ \pi^0) &=& -i \frac{f^*_{dc} f_{du}}{8\sqrt{2}m^2_H} \left(
 1+\frac{1}{N_c}\right)f_\pi f_{0}(0) m^2_D\,, \non\\
 A(\pi^+ \phi) &=& -i \frac{f^*_{dc} f_{du}}{8m^2_H} \left(
 1+\frac{1}{N_c}\right)f_\phi f_{+}(0) p_1\cdot \ve_\phi \,.
 \ed
Here, we have taken the approximation of $m^2_\pi\approx 0$ and set
$N_c =3$. As a result, the branching ratios (BRs) are known as
 \be
B(D^+\to \pi^+ \pi^0) &=& \frac{\tau_D m^3_D}{2^{11} \pi}\left(
 1+\frac{1}{N_c}\right)^2 f^2_\pi f^2_0(0) \left| \frac{f^*_{dc}
 f_{du}}{m^2_H}\right|^2\,,\non\\
B(D^+\to \pi^+ \phi) &=& \frac{\tau_D m^3_D}{2^{12} \pi}\left(
 1+\frac{1}{N_c}\right)^2 f^2_\phi f^2_+(0) \left| \frac{f^*_{sc}
 f_{su}}{m^2_H}\right|^2\,.
 \ed
For simplicity, we use the central values of the data to obtain the
upper limits of the parameters. With $f_\pi =0.13$ GeV,
$f_\phi=0.237$ GeV and $f_{+}(0)=f_0(0)=0.624$ \cite{formfactor}, we
get
 \be
\left| \frac{f^*_{dc}
 f_{du}}{m^2_H}\right|^2 &<& 1.7\times 10^{-10} \rm GeV^{-4}\,,\non\\
\left| \frac{f^*_{sc}
 f_{su}}{m^2_H}\right|^2 &<& 1.5\times 10^{-9} \rm GeV^{-4}\,.
 \ed
By comparing with Eq.~(\ref{eq:xD_limit}), we see clearly that unless
there exist strong cancelations among the free parameters, the
constraints from the $D^0-\bar D^0$ mixing are much stronger than those
from D decays.

By examining  Fig.~\ref{fig:H6_loop}, it is easy to find that
down-type quarks involve in the internal loop. In other words, the
$K^0-\bar K^0$ mixing, denoted by $\Delta m_K$, might give a strict
constraint on the parameters. To understand the influence of $\Delta
m_K$, by the similar calculations on $\Delta m_D$, the formula is
given by
 \be
\Delta m_K &=& 2Re\langle \bar K^0 | H_{\Delta S=2} |  K^0 \rangle=
\frac{19f^2_K m_K B_K}{768 \pi^2}\left(\frac{ReC^2_K}{m^2_H} \right)
\label{eq:deltamK}
 \ed
with $C_K=\sum_{j=u,c,t} f_{sj}f^*_{dj}$, where we have set $m_t\ll
m_{H}$. Clearly, although some parameters such as $f_{sc}f^*_{dc}$
and $f_{su}f^*_{du}$ appear in both $\Delta m_D$ and $\Delta m_K$,
in general the $C_K$ and $C_D$ are different parameters. Adopting
this viewpoint, the constraint from $\Delta m_K$ might not have an
influence on the constraint from $\Delta m_D$. Nevertheless, in some
special case, $C_K$ and $C_D$ are strongly correlated. Therefore, it
is interesting to survey both constraints in a little bit of detail.
Hence, according to Eqs.~(\ref{eq:deltamD}) and (\ref{eq:deltamK}),
we get
 \be
\frac{\Delta m_K}{\Delta m_D} \approx \frac{f^2_K m_K B_K}{f^2_D m_D
B_D} \frac{|C_K|^2}{|C_D|^2}\approx 0.14 \frac{|C_K|^2}{|C_D|^2}\,.
 \ed
With the data of $\Delta m_D=x_D\Gamma_D\sim 0.008\Gamma_D \approx
1.3\times 10^{-14}$ GeV and $\Delta m_K=3.483 \times 10^{-15}$ GeV
\cite{PDG08}, the constrained relation from D and K systems is
 \be
 \frac{|C_K|}{|C_D|}\approx 1.4.
 \ed
Clearly, the result implies that when $|C_K|$ and  $|C_D|$ are not
regarded as independent parameters, both $\Delta m_K$ and $\Delta
m_D$ will give similar constraints on the free parameters.

\underline {2. \textit{diquark (6,1,4/3)} }

\begin{figure}[htbp]
\includegraphics*[width=4.0 in]{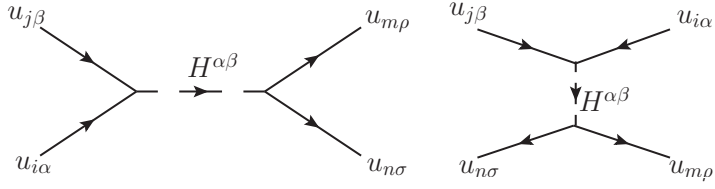}
\caption{Diquark tree diagrams for the $D^0-\bar D^0$ mixing.}
 \label{fig:tree_H6}
\end{figure}

Next, we consider the contributions of the diquark (6,1,4/3), where the
couplings to up-type quarks are written by
 \be
{\cal L}_{H_{\bf 6}} = f_{ij} u^{T}_{i\alpha} C P_R u_{j\beta} H^{
\dagger \alpha\beta}_{\bf 6} +h.c. \label{eq:L_6143}
 \ed
Here, we adopt the same notations as those used for (6,1,1/3).
However, one can find that the behavior of couplings $f_{ij}$ is
symmetric in flavors. To illustrate the result, the derivation is
given as follows:
 \be
f_{ij} u^{T}_{i\alpha} C P_R u_{j\beta} H^{\dagger\alpha\beta}_{\bf
6} &=& - f_{ij} u^T_{j\beta} P_R C^T u_{i\alpha}
H^{\dagger\alpha\beta}_{\bf 6}
 \non\\
 = f_{ji} u^T_{i\beta}  C P_R u_{j\alpha} H^{\dagger \alpha\beta}_{\bf 6}&=&
f_{ji} u^T_{i\alpha}  C P_R u_{j\beta} H^{\dagger \alpha\beta}_{\bf
6},
 \ed
where we have used $H^{\alpha\beta}_{\bf 6}= H^{\beta\alpha}_{\bf
6}$ in the last equality. Clearly, we get $f_{ij}=f_{ji}$. The
flavor diagrams for D decays are displayed in
Fig.~\ref{fig:tree_H6}.
 To produce the $\Delta C=2$ process, the flavor
indices in Fig.~\ref{fig:tree_H6} are chosen to be $i=j=c$ and
$m=n=u$.
With Wick contractions, the associated
four-fermion interactions are obtained as
 \be
{\cal H}_{\Delta C=2}= - \frac{f_{cc} f^*_{uu} }{4m^2_H} \left[\bar
u\ga_\mu P_R c \bar u\ga^\mu P_R c + \bar u_\beta \ga_\mu P_R
c_{\alpha} \bar u_{\alpha} \ga^\mu P_R c_{\beta} \right]\,.
 \ed
Using the transition matrix element of Eq.~(\ref{eq:FDD}), we find
 \be
  \langle \bar D^0| {\cal H}_{\Delta C=2} |
 D^0\rangle = -\frac{7}{24}
 f^2_D m^2_D B_D
\frac{f_{cc}f^*_{uu} }{m^2_H}\,.
 \ed
Consequently, the mixing parameter of the $D^0-\bar D^0$ mixing is
 \be
 x_D&=&- \bar C_{D} \frac{Re(f_{cc}f^*_{uu}) }{m^2_H}\,, \non \\
 \bar C_{D}&=&\frac{7}{24}\tau_D
f^2_D m_D B_D\,.
 \ed
 From the values taken previously, the bound on the free parameters is
found as
 \be
 \left|\frac{Re(f_{cc}f^*_{uu}) }{m^2_H}\right|= 7.2 \times 10^{-11} x_D \rm GeV^{-2}\,.
 \ed
If we adopt $m_H\sim 1$ TeV and $x_D\sim 8\times 10^{-3}$, the
constraint on the parameter is $|Re(f_{cc} f^*_{uu})|\sim 5.76
\times 10^{-7}$. In particular,
for $f_{cc} \sim f_{uu}$
one gets $|f_{cc}|\sim |f_{uu}|\sim 7.6\times 10^{-4}$.

At the first sight, it seems that the small couplings are
fine-tuned. However, the smallness in fact could be related to the
suppressed flavor mixings. To demonstrate the conjecture, we propose
that before the electroweak symmetry breaking, the scalar diquark only
couples to one flavor, such as the  top quark. Accordingly, the interaction
is set to be
 \be
{\cal L}_{ttH_{\bf 6}} = f_{t} t^{T}_{\alpha} C P_R t_{\beta} H^{
\dagger \alpha\beta}_{\bf 6} +h.c.\label{eq:L_ttH}
 \ed
After the symmetry breaking,  to diagonalize the up-quark mass matrix,
we introduce unitary matrices $V_L$ and $V_R$ so that the physical
and weak eigenstates are related by $u_{R(L)} = V_{R(L)}
u^w_{R(L)}$. Using the relation, the couplings in
Eq.~(\ref{eq:L_6143}) can be related to $f_t$ of
Eq.~(\ref{eq:L_ttH}) and $V_R$ by
 \be
 f_{ij} = f_t \left( V_{Ri3} V_{Rj3} \right)^*\,.
 \ed
If we take $V_{R23}\sim \lambda^2$ and $V_{R13}\sim \lambda^3$ with
$\lambda\sim 0.22$, then $f_{cc}f^*_{uu}= f^2_t V^{*2}_{R23}
V^2_{R13}\sim f^2_t \lambda^{10}=2.7 f^2_t\times 10^{-7}$. By
choosing suitable value for $f_t$, the proposed scenario in the
analysis can naturally fit the result $|Re(f_{cc} f^*_{uu})|\sim
5.76 \times 10^{-7}$ from the $D^0-\bar D^0$ mixing. Intriguingly,
the proposed model is similar to the case studied by the authors in
Ref.~\cite{MOY_PRD77} which is based on the $SU(2)_L\times SU(2)_R
\times SU(4)_C$ symmetry. Moreover, the detailed studies on the
production of colored sextet scalar at Colliders could be referred
to Refs.~\cite{MOY_PRD77,CKRW}.

In summary, we have investigated the contributions of the scalar
diquarks to the $D^0-\bar D^0$ oscillation. We have shown that the
scalar diquark of (6, 1, 1/3) can generate the $\Delta C=2$ process
through box diagrams. By comparing with Cabibbo- and
color-suppressed  $D$ decays, the most strict limit on the free
parameters  arises from $\Delta m_D$. We also show that the
constraints of $\Delta m_D$ and $\Delta m_K$ are comparable. For the
diquark of (6, 1, 4/3), we have found that the $\Delta C=2$ process
can be induced at tree level. Although $\Delta m_D$ gives a very
strong constraint on the free parameters, for the model with the
diquark only coupling to the top-quark, the resultant small
couplings can be ascribed to the small elements of
the flavor mixing.\\

 \noindent {\bf Acknowledgments}

This work is supported by the National Science Council of R.O.C.
under Grant No: NSC-97-2112-M-006-001-MY3.

\end{document}